\documentclass[twoside,11pt]{article}

\usepackage{graphicx}
\usepackage{amssymb}
\usepackage{amsmath}
\usepackage{amsfonts}
\usepackage{wasysym}
\usepackage{subfigure}

\begin{document}           % End of preamble and beginning of text.

\title{Universe on Extremely Small Spacetime Scales: Quantum Geometrodynamical Approach}
\author{V.E. Kuzmichev, V.V. Kuzmichev\\[0.5cm]
\itshape Bogolyubov Institute for Theoretical Physics,\\
\itshape National Academy of Sciences of Ukraine, Kiev, 03680 Ukraine}

\date{}

\maketitle

\begin{abstract}
The semi-classical approach to the quantum geometrodynamical model is used for
the description of the properties of the universe on extremely small spacetime scales.
Quantum theory for a homogeneous, isotropic and closed universe is constructed 
on the basis of a Hamiltonian formalism with the use of material reference system 
as a dynamical system defined by macroscopic relativistic matter.
Under this approach the equations of the model are reduced to the form of the
Einstein-type equations in which the matter energy density has two components of quantum
nature, which behave as antigravitating fluids. The first component does not vanish
in the limit $\hbar \rightarrow 0$ and can be associated with dark energy. The second
component is described by extremely rigid equation of state and goes to zero after
the transition to large spacetime scales. On small spacetime scales this quantum 
correction turns out to be significant. It determines the geometry of the universe
near the initial cosmological singularity point. This geometry
is conformal to a unit four-sphere embedded in a five-dimensional
Euclidean flat space. During the consequent expansion of the 
universe, when reaching the post-Planck era, 
the geometry of the universe changes into the geometry conformal to 
a unit four-hyperboloid in a five-dimensional Lorentz-signatured flat space.
This agrees with the 
hypothesis about the possible change of geometry after the origin of 
expanding universe from the region near the initial singularity point.
The origin of the universe can be interpreted as a quantum transition of the
system from the region in a phase space forbidden for classical motion, but where
a trajectory in imaginary time exists, into the region, where the equations of motion
have the solution which describes the evolution of the universe in real time.
The calculated transition amplitude appears to be exponentially high.
Near the
boundary between two regions, from the side of real time, the universe undergoes
almost an exponential expansion which passes smoothly into the expansion under the
action of radiation dominating over matter which is described by the standard
cosmological model. 
The mechanism of a shrinkage of the region forbidden for the classical motion
to the point of the initial cosmological singularity is described.
\end{abstract}


\begin{center}
      \textbf{1. Introduction}\\[0.3cm]
\end{center}

It is accepted that the present-day Universe as a whole can be considered as
a cosmological system described by the standard
model based on general relativity. According to the Standard
Big Bang Model \cite{OP,KolT}, the early Universe was very hot and dense. In order to
describe that era one must take into account that in the course of its 
evolution the Universe has passed through a stage with quantum degrees of freedom
of the gravitational and matter fields 
before turning into the cosmological system, whose properties are described well by 
general relativity. It means that a consistent description of the Universe as a 
nonstationary cosmological system should be based on quantum general relativity in the 
form admitting the passage to general  relativity in semi-classical limit 
$\hbar \rightarrow 0$ \cite{Ish,Kie}.

A consistent quantum theory of gravity, in principle, can be constructed on the 
basis of the Hamiltonian formalism with the application of the canonical 
quantization method.
The first problem on this way is to choose generalized 
variables. In a straightforward manner (see, e.g. Refs. \cite{And,Whe,DeW}) one may choose
metric tensor components and matter fields as such variables. But the functional
equations obtained in this approach prove to be insufficiently
suitable for specific problems of quantum theory of gravity and cosmology. These
equations do not contain a time variable in an explicit form. This, in turn, 
gives rise
to the problem of interpretation of the state vector of the universe (see e.g.
discussion in Ref. \cite{KolT} and references therein). A cause of the failure
can be easily understood with the help of Dirac's constraint system theory 
\cite{Dir}. It is found that the structure of constraints in general relativity is 
such that variables which correspond to true dynamical degrees of freedom cannot be 
singled out from canonical variables of geometrodynamics. This difficulty is 
stipulated by an absence of predetermined way to identify spacetime events in 
generally covariant theory \cite{KuchT}

One of the possible versions of a theory
with a well-defined time variable is proposed in Refs. \cite{KK1,KK2} in the case
of homogeneous, isotropic and closed universe.
The universe is supposed to be filled with a uniform scalar field which stands for
the primordial matter\footnote{Since we deal with the quantum theory we should describe
the matter content of the universe by some fundamental Lagrangians of fields.} and 
(macroscopic) relativistic matter associated with \textit{material} reference system.
The importance of material reference systems as dynamical systems for quantum
gravity was indicated by DeWitt \cite{DeW,KuchT,BM}. Such an approach implies that the 
reference system is related to some true physical medium.

As calculations have demonstrated \cite{KK1,KK2}, the equations of the
quantum model may be reduced to the form in which the matter energy density in the
universe has a component which is a condensate of massive quanta of a 
scalar field. Under the semi-classical description this component behaves as
an antigravitating fluid. Such a property has a quantum nature and it is connected
with the fact that the states with all possible masses of a condensate
contribute to the state vector of the quantum universe. If one discards the 
corresponding quantum corrections, the quantum fluid degenerates into a dust, i.e.
matter component of the energy density commonly believed to make a dominant 
contribution to the mass-energy of ordinary matter in the present Universe in 
the standard cosmological model. Let us note that the presence of a condensate
in the universe, as well as the availability of a dust representing an extreme state of
a condensate, is not presupposed in the initial Lagrangian of the theory.
If one supposes that the properties of our Universe are
described in an adequate manner by such a quantum theory, an antigravitating 
condensate being found out can be associated with dark energy \cite{KK2}.
Assuming that particles of a condensate can decay to baryons, leptons (or to their
antiparticles) and particles of dark matter, one can describe the percentage of
baryons, dark matter and dark energy observed in the Universe \cite{KK3}.

In semi-classical limit the negative pressure fluid arises as a remnant of the early
quantum era. This antigravitating component of the energy density does not
vanish in the limit $\hbar \rightarrow 0$. In addition to this component, the
stress-energy tensor contains the term vanishing after the transition to
general relativity, i.e. to large spacetime scales. However, on small spacetime
scales quantum corrections  $\sim \hbar$ turn out to be significant. As it is shown
in this paper, the effects stipulated by these corrections determine the 
equation of state of matter and geometry near the initial 
cosmological singularity point. They define a boundary condition that should be imposed
on the state vector in the origin so that a nucleation of the universe from the 
initial cosmological singularity point becomes possible.

In Sect.~2 the basic principles and equations for classical and quantum models 
are formulated. Sect.~3 is devoted to reduction of equations of quantum theory
for the case of specific scalar field model. In Sect.~4 the quantum effects are 
investigated in semi-classical approximation. The special emphasis is put on 
studying quantum effects on sub-Planck scales. Sect.~5 presents some concluding remarks.

In this paper we use the modified Planck system of units. The
$l_{P} = $ $\sqrt{2 G \hbar /(3 \pi c^{3})}$ is taken as a unit of length,
the $\rho_{P} = 3 c^{4} /(8 \pi G l_{P}^{2})$ is a unit of energy density and so on.
All relations are written for dimensionless values.

\begin{center}
      \textbf{2. Equations of motion }\\[0.3cm]
\end{center}

\textbf{2.1. Classical model.}
Let us consider the homogeneous, isotropic and closed universe which is described
by the Robertson-Walker metric
\begin{equation}\label{1}
     ds^{2} = d\tau^{2} - a^{2} d\Omega_{3}^{2},
\end{equation}
where $\tau$ is the proper time, $a$ is the cosmic scale factor, $d\Omega_{3}^{2}$ is a 
line element on a unit three-sphere. It is convenient to pass to a new time variable 
$\eta$,
\begin{equation}\label{2}
    d\tau = a N d\eta,
\end{equation}
where $N$ is the lapse function that specifies the time reference scale.

We assume that the universe is originally filled with a uniform 
scalar field $\phi$ and a perfect fluid. The latter as a macroscopic medium defines 
so called material reference system \cite{KuchT,KK1,BM, KK4, T}. 
In the model of the universe under consideration the action has the form
\begin{equation}\label{3}
    S = \int d\eta \left\{\pi_{a}\,\frac{da}{d\eta} + \pi_{\phi}\,\frac{d\phi}{d\eta} + 
         \pi_{\Theta}\,\frac{d\Theta}{d\eta} + 
\pi_{\tilde{\lambda}}\,\frac{d\tilde{\lambda}}{d\eta} - H \right\},
\end{equation}
where  $\pi_{a},\, \pi_{\phi},\, \pi_{\Theta},\, \pi_{\tilde{\lambda}}$ are the momenta 
canonically conjugate with the variables $a,\, \phi,\, \Theta,\, \tilde{\lambda}$,
\begin{eqnarray}\label{4}
    H = \frac{N}{2}\left\{-\,\pi_{a}^{2} - a^{2} + a^{4} [\rho_{\phi} + \rho]\right\} 
+ \lambda_{1}\left\{\pi_{\Theta} - \frac{1}{2}\,a^{3} \rho_{0} s\right\}
+ \lambda_{2}\left\{\pi_{\tilde{\lambda}} + \frac{1}{2}\,a^{3} \rho_{0} \right\}
\end{eqnarray}    
is the Hamiltonian, 
\begin{equation}\label{5}
    \rho_{\phi} = \frac{2}{a^{6}}\,\pi_{\phi}^{2} + V(\phi)
\end{equation}
is  the energy density of a scalar field with the potential $V(\phi)$, 
$\rho = \rho(\rho_{0}, s)$ is the energy density of a perfect fluid which is
a function of the density of the rest mass $\rho_{0}$ and the specific entropy $s$. 
The $\Theta$ is the thermasy (potential for the temperature, $T = \Theta_{,\,\nu} 
U^{\nu}$). The $\tilde{\lambda}$ is 
the potential for the specific free energy $f$ taken with an inverse sign, $f = 
-\,\tilde{\lambda}_{,\,\nu} U^{\nu}$. The $U^{\nu}$ is the four-velocity.
The momenta $\pi_{\rho_{0}}$ and $\pi_{s}$
conjugate with the variables $\rho_{0}$ and $s$ vanish identically,
\begin{equation}\label{6}
    \pi_{\rho_{0}} = 0, \qquad \pi_{s} = 0.
\end{equation}
The Hamiltonian (\ref{4}) of such a system has the form of a linear 
combination of constraints and weakly vanishes (in Dirac's sence 
\cite{Dir}),
\begin{equation}\label{7}
    H \approx 0,
\end{equation}
where the sign $\approx$ means that Poisson brackets must all be worked out 
before the use of the constraint equations.
The $N$, $\lambda_{1}$, and $\lambda_{2}$ are Lagrange multipliers.
The variation of the action (\ref{3}) with respect to them leads to three constraint 
equations
\begin{equation}\label{8}
    -\,\pi_{a}^{2} - a^{2} + a^{4} [\rho_{\phi} + \rho] \approx 0, \quad
    \pi_{\Theta} - \frac{1}{2}\,a^{3} \rho_{0} s \approx 0, \quad
     \pi_{\tilde{\lambda}} + \frac{1}{2}\,a^{3} \rho_{0} \approx 0.
\end{equation}

From the conservation of these constraints in time it follows that
the conservation laws hold,
\begin{equation}\label{9}
    E_{0} \equiv a^{3} \rho_{0} = \mbox{const}, \qquad s =  \mbox{const},
\end{equation}
where the first relation describes the conservation law of a
macroscopic value which characterizes the number of particles of a perfect fluid, the 
second equation represents the conservation of the specific entropy.
Taking into account these conservation laws 
and the equations (\ref{6}) one can discard degrees of freedom corresponding to the 
variables $\rho_{0}$ and $s$, and convert the second-class constraints into
first-class constraints \cite{KK1} in accordance with Dirac's proposal.

The equation of motion for the classical dynamical variable $\mathcal{O} = 
\mathcal{O}(a, \phi, \pi_{a}, \pi_{\phi}, \dots )$ has the form
\begin{equation}\label{10}
    \frac{d\mathcal{O}}{d\eta} \approx \{\mathcal{O}, H\},
\end{equation}
where $H$ is the Hamiltonian (\ref{4}), $\{.,.\}$ are Poisson brackets. 

\textbf{2.2. Quantum model.}
In quantum theory first-class constraint equations (\ref{8}) become constraints on the
state vector $\Psi$. Passing from classical variables to corresponding operators and 
using the conservation laws (\ref{9}) we obtain three equations
\begin{eqnarray}\label{11}
   \left\{-\,\partial^{2}_{a} + a^{2} - a^{4} [\rho_{\phi} + \rho] \right\} \Psi = 0, 
\nonumber \\
   \left\{ -\,i \partial_{\Theta} - \frac{1}{2}\,E_{0} s \right\} \Psi = 0, \quad
   \left\{ -\,i \partial_{\tilde{\lambda}} + \frac{1}{2}\,E_{0} \right\} \Psi = 0.
\end{eqnarray}
It is convenient to pass from the generalized variables $\Theta$ and 
$\tilde{\lambda}$ to the non-coordinate co-frame
\begin{eqnarray}\label{12}
  h\, d\tau = s\,d\Theta\, -\,d\widetilde{\lambda},\qquad
  h\, dy = s\,d\Theta\, +\,d\tilde{\lambda},
\end{eqnarray}
where $h = \frac{\rho + p}{\rho_{0}}$ is the specific enthalpy which plays the role
of inertial mass, $p$ is the pressure, $\tau$ is proper time in every 
point of space, and $y$ is supplementary variable. The corresponding derivatives commute 
between themselves, $\left[\partial_{\tau},\,\partial_{y}\right] = 0$.

From the first equation in the set (\ref{11}) it follows that it is convenient to choose
a perfect fluid in the form of relativistic matter. Introducing the value 
\begin{equation}\label{13}
    E \equiv a^{4} \rho = \mbox{const},
\end{equation} 
we come to the equations which describe the quantum universe  \cite{KK1}
\begin{equation}\label{14}
    \left\{-\, i\,\partial_{\tau_{c}} - \frac{1}{2}\, E_{0} \right\} \Psi = 0,
 \qquad \partial_{y} \Psi = 0,
\end{equation}
\begin{equation}\label{15}
    \left\{-\,\partial^{2}_{a} +  a^{2} - 2 a \hat{H}_{\phi} - E\right\} \Psi = 0,
\end{equation}
where  $\tau_{c}$ is the time variable connected
with the proper time $\tau$ by the differential relation  $d\tau_{c} = h\, d\tau$,
\begin{equation}\label{16}
    \hat{H}_{\phi} = \frac{1}{2}\,a^{3}\left[-\,\frac{2}{a^{6}}\,\partial_{\phi}^{2} + 
V(\phi)\right]
\end{equation}
is the operator of mass-energy of a scalar field in a comoving volume 
$\frac{1}{2}\, a^{3}$. From the equations (\ref{14}) it follows that 
 $\Psi$ does not depend on the variable $y$. The first equation of the set (\ref{14})
 has a particular solution in the form
\begin{equation}\label{17}
    \Psi = e^{\,\frac{i}{2}\,E \bar{\tau}} |\psi \rangle,
\end{equation}
where $\bar{\tau} = \frac{E_{0}}{E} \tau_{c}$ is the rescaled time variable.
The state vector $|\psi\rangle$ is defined in the space of two variables  $a$ and 
$\phi$, and determined by the equation
\begin{equation}\label{18}
    \left(-\,\partial^{2}_{a} +  a^{2} - 2 a \hat{H}_{\phi} - E\right)|\psi\rangle = 
0.
\end{equation}

This equation describes the state of the universe with definite value of the
parameter $E$. Eqs. (\ref{14}) and (\ref{15}) with regard for Eq. (\ref{17})
are equivalent to the Schr\"{o}dinger-type equation obtained in Ref. \cite{KK4}
by means of a coordinate condition introduced to specify a reference system.

The vector $|\psi \rangle$ represents the dynamical state of the universe 
at some instant of time $\eta_{0}$ which is
connected with time $\bar{\tau}$ by the equation $\bar{\tau} = \frac{4}{3}\int 
^{\eta_{0}}Nd\eta$. Supposing that the vector $|\psi \rangle$ is normalized to unity
\cite{KK1} and considering it as immovable vector of the Heisenberg representation one 
can describe the motion of the quantum universe by the equation
\begin{equation}\label{19}
      \langle \psi|\frac{1}{N}\, \frac{d}{d\eta} \hat{\mathcal{O}}|\psi\rangle =
    \frac{1}{N}\, \frac{d}{d\eta}\langle \psi|\hat{\mathcal{O}}|\psi\rangle =
     \frac{1}{i} \, \langle \psi|[\hat{\mathcal{O}},\frac{1}{N} \hat{H}]|\psi\rangle, 
\end{equation}
where $[.,.]$ is a commutator, and $\hat{H}$ is determined by the expression
(\ref{4}), in which all dynamical variables are substituted with operators. The 
observable $\hat{\mathcal{O}}$  corresponds to the classical dynamical variable 
$\mathcal{O}$. For  $\hat{\mathcal{O}} = a$ we obtain
\begin{equation}\label{20}
        \langle \psi|-i\partial_{a}|\psi\rangle  = 
       - \langle  \psi|a \frac{da}{d\tau}|\psi\rangle.
\end{equation}
In the classical theory  corresponding momentum has the form
\begin{equation}\label{21}
    \pi_{a} = \partial_{a} S = - a \frac{da}{d\tau}  \equiv - a\dot{a},
\end{equation}
where $S$ is the action. For $\hat{\mathcal{O}} = - i\partial_{a}$ we find
\begin{equation}\label{22}
    \langle \psi|- i \frac{1}{N} \frac{d}{d\eta} \partial_{a}|\psi\rangle = 
  \langle \psi|a - \frac{2}{a^{3}}\, \partial_{\phi}^{2} - 2 a^{3} V(\phi)|\psi\rangle .
\end{equation}

\begin{center}
      \textbf{3. Scalar field model}\\[0.3cm]
\end{center}

Eq. (\ref{18}) can be integrated with respect to $\phi$, if one determines the
form of the potential $V(\phi)$. 
As in Ref. \cite{KK2} we consider the solution of Eq. (\ref{18}) in the era when the 
field $\phi$ oscillates with a small amplitude near the minimum of its 
potential at the point $\phi = \sigma$. 
Then $V(\phi)$ can be approximated by the expression 
\begin{equation}\label{23}
    V(\phi) = \rho_{\sigma} + \frac{{m}_{\sigma}^{2}}{2}\, (\phi - \sigma)^{2},
\end{equation}
where $\rho_{\sigma} = V(\sigma)$, ${m}_{\sigma}^{2} = 
[d^{2}V(\phi)/d\phi^{2}]_{\sigma} > 0$.
If $\phi = \sigma$ is the point of absolute minimum, then $\rho_{\sigma} = 0$ and
the state $\sigma$ corresponds to the true vacuum of a primordial scalar field, 
while the state with $\rho_{\sigma} \neq 0$ matches with the false vacuum \cite{Col}. 

Introducing the new variable
\begin{equation}\label{24}
    x = \left(\frac{m_{\sigma} a^{3}}2{}\right)^{1/2} (\phi - \sigma),
\end{equation}
which describes a deviation of the field $\phi$ from its equilibrium state,
and defining the functions of harmonic oscillator $\langle x |u_{k}\rangle$
as solutions of the equation
\begin{equation} \label{25}
   \left(- \partial_{x}^{2} + x^{2}\right)|u_{k}\rangle = (2k + 1) |u_{k}\rangle, 
\end{equation}
where $k = 0,\,1,\,2,\,... $ is a number of state of the oscillator,
we find
\begin{equation}\label{26}
    \hat{H}_{\phi} |u_{k}\rangle = \left(M_{k} + \frac{1}{2}\,a^{3} 
\rho_{\sigma}\right)|u_{k}\rangle,
\end{equation}
where the quantity
\begin{equation}\label{27}
     M_{k} = m_{\sigma} \left(k + \frac{1}{2}\right)
\end{equation}
can be interpreted as an amount of matter-energy (or mass) in the universe 
related to a scalar field.
This energy is represented in the form of a sum of excitation quanta of 
the spatially coherent oscillations of the field $\phi$ about the equilibrium state
$\sigma$, $k$ is the number of these excitation quanta. The mentioned
oscillations correspond to a condensate of zero-momentum $\phi$ quanta with the mass 
$m_{\sigma}$.

We shall look for the solution of Eq. (\ref{18}) in the form of the superposition of
the states with different masses $M_{k}$, 
\begin{equation}\label{28}
    |\psi\rangle = \sum_{k}\, |f_{k}\rangle |u_{k}\rangle.
\end{equation}
Using orthonormality of the  $|u_{k}\rangle$ we obtain the equation for
the vector $|f_{k}\rangle$
\begin{equation}\label{29}
     \left(- \partial_{a}^{2} + U_{k} - E\right) |f_{k}\rangle = 0,
\end{equation}
where
\begin{equation}\label{30}
    U_{k} = a^{2} - 2 a M_{k} - a^{4} \rho_{\sigma}
\end{equation}
is the effective potential. In the case $\rho_{\sigma} = 0$ this equation is exactly 
integrable \cite{KK1}. The corresponding eigenvalue is equal to
\begin{equation}\label{31}
    E \equiv E_{n,k} = 2n + 1 - M_{k}^{2}, 
\end{equation}
where $n = 0,\,1,\,2,\,... $ is a number of state of the quantum universe with the mass 
$M_{k}$ in the potential well (\ref{30}). The vectors $|f_{k}\rangle$ and 
$|f_{k'}\rangle$ at $k \neq k'$ are, generally speaking, 
nonorthogonal between themselves.
So that the transition probability $w(n,k \rightarrow n',k') = |\langle 
f_{k'}|f_{k}\rangle | ^{2}$ is nonzero. For example, 
the  probability of transition of the universe 
from the ground (vacuum) state  $n = 0$ to any other state obeys the Poisson distribution
\begin{equation}\label{32}
    w(0,k \rightarrow n',k') = \frac{\langle n' \rangle ^{n'}}{n'!}\,
           e^{-\langle n' \rangle}, 
\end{equation}
where $\langle n' \rangle = \frac{1}{2}\, (M_{k'} - M_{k})^{2}$ is the mean value of the
quantum number $n'$.

Substituting Eq. (\ref{28}) into Eq. (\ref{22}) we obtain
\begin{equation}\label{33}
    \langle f_{k}|- \frac{i}{N}\,\frac{d}{d\eta} \partial_{a}| f_{k}\rangle =
     \langle f_{k}|a -2 a^{3}\rho_{\sigma} - 4M_{k}| f_{k}\rangle + \Delta_{k},
\end{equation}
where 
\begin{equation}\label{34}
   \Delta_{k} = - 3 m_{\sigma} \langle f_{k}| \sum _{k'} \langle u_{k}| 
                 \partial_{x}^{2}|u_{k'}\rangle | f_{k'}\rangle .
\end{equation}
Here $k'$ takes the values $k$ and $k\pm 2$. This term describes the component of the 
pressure of the condensate stipulated by the motion of $\phi$ quanta in phase space 
with the momentum $-i\partial_{x}$. Using Eq. (\ref{25}) we find
\begin{eqnarray}\label{35}
    \Delta_{k}   = 3 M_{k} \langle f_{k}|f_{k}\rangle - 
    \frac{3}{2}\sqrt{\left(M_{k}  + \frac{3}{2}\,m_{\sigma}\right)\left(M_{k} + 
      m_{\sigma}\right)}\langle f_{k}|f_{k+2}\rangle  \\ -
      \frac{3}{2}\sqrt{\left(M_{k} - \frac{3}{2}\,m_{\sigma}\right)\left(M_{k} - 
      m_{\sigma}\right)}\langle f_{k}|f_{k-2}\rangle \nonumber
.
\end{eqnarray} 
In the case $k \gg 1$ the masses $M_{k\pm 2} \simeq  M_{k} \gg \frac{1}{2}\,m_{\sigma}$
and according to Eq. (\ref{29}) the vectors $|f_{k \pm 2}\rangle \simeq |f_{k}\rangle $.
Then
\begin{equation}\label{36}
     \Delta_{k} = 0 \qquad \mbox{at} \qquad k \gg 1.
\end{equation}
It means that the contributions into the sum with respect to $k'$ in Eq. (\ref{34}) from
the different $k$-states of the universe are mutually canceled. As a result the pressure
of a condensate is determined only by its quantum properties (see bellow Eq. (\ref{44})). 
We note that
if one discards the contributions from the transition amplitudes $\langle f_{k}|f_{k \pm
 2}\rangle $, a condensate turns into an aggregate of separate macroscopic bodies with 
zero pressure (dust) \cite{KK2}. The existence of this limit argues in favour
of reliability of this quantum model.

\begin{center}
      \textbf{4. Semi-classical approach }\\[0.3cm]
\end{center}

\textbf{4.1. Einstein-type equations.}
In order to give the physical meaning to the different quantities emergent in this theory
we reduce Eqs (\ref{29}) and (\ref{33}) to the form of the Einstein
equations. With that end in view we choose the vector $|f_{k}\rangle $ in the form
\begin{equation}\label{37}
   \langle a|f_{k}\rangle  = \frac{\mbox{const}}{\sqrt{\partial_{a} S(a)}}\, 
                         e^{i S(a)},
\end{equation}
where $S$ is unknown function of $a$ (the index $k$ we omit here and below). 
Substituting Eq. (\ref{37}) into (\ref{29}) and (\ref{33}) with regard to (\ref{36}) 
we obtain
\begin{equation}\label{38}
   \frac{1}{a^{4}} \left(\partial_{a} S \right)^{2} - \rho_{m}  - \rho_{u} + 
            \frac{1}{a^{2}} = 0,
\end{equation}
\begin{equation}\label{39}
   \frac{1}{a^{2}}\,\frac{d}{d \tau}\,\left(\partial_{a} S \right) + 
     \frac{1}{2}\,\left(\rho_{m} - 3 p_{m} \right) + \widetilde{\rho}_{u} -
      \frac{1}{a^{2}} = 0, 
\end{equation}
where
\begin{equation}\label{40}
    \rho_{u} = \frac{1}{a^{4}}\,\left\{\frac{3}{4}\, \left(\frac{\partial_{a}^{2} 
            S}{\partial_{a}S}\right)^{2} - \frac{1}{2}\,\frac{ \partial_{a}^{3} 
            S}{\partial_{a}S}\right\}
\end{equation}
and
\begin{equation}\label{41}
   \widetilde{\rho}_{u} = \frac{i}{2 a^{2}}\,\frac{d}{d \tau}\,
               \left(\frac{\partial_{a}^{2} S}{\partial_{a}S} 
\right)
\end{equation}
are the quantum corrections to the stress-energy tensor, $\rho_{u}\sim \hbar^{2}$  and 
$\widetilde{\rho}_{u} \sim \hbar$ (in ordinary units \cite{KK2}),
\begin{equation}\label{42}
    \rho_{m} = \rho_{k} +\rho_{\sigma} + \rho, \qquad 
     p_{m} = p_{k} + p_{\sigma} + p
\end{equation}
are the energy density and the isotropic pressure as the sums of the components,
\begin{equation}\label{43}
    \rho_{k} = \frac{2 M_{k}}{a^{3}}, \qquad  \rho_{\sigma} \equiv V(\sigma) \equiv 
      \frac{\Lambda}{3}, \qquad \rho = \frac{E}{a^{4}},
\end{equation}
$\Lambda$ is the cosmological constant. The equations of state are
\begin{equation}\label{44}
   p_{k} = - \rho_{k}, \qquad  p_{\sigma} = - \rho_{\sigma},\qquad 
       p = \frac{1}{3}\,\rho. 
\end{equation}
The equations of state for the vacuum component $\rho_{\sigma} = \mbox{const}$ and 
relativistic matter $\rho$ are dictated by the formulation of the problem. 
The vacuum-type equation of state of a condensate with the density $\rho_{k}$, which
does not remain constant throughout the evolution of the universe, but decreases 
according to a power law with the increase of $a$, follows from the 
condition of consistency of Eqs (\ref{38}) and (\ref{39}).

From Eqs (\ref{42}) - (\ref{44}) we can conclude that a condensate behaves as an 
anti-gravitating medium. Its anti-gravitating effect has a purely quantum nature. 
Its appearance is determined by the fact that the state vector of the universe
(\ref{28}) is a superposition of quantum states with all possible values of the quantum 
number $k$. This component of energy density does not vanish in the limit $\hbar \rightarrow 0$.

In the classical limit $(\hbar = 0)$ the terms $\rho_{u}$ and $\widetilde{\rho}_{u}$
can be discarded and Eqs (\ref{38}) and (\ref{39}) reduce to the Einstein equations
which predict an accelerating expansion of the universe in the era with 
$\rho_{k} > \frac{2}{3}\,\rho$, even if $\Lambda = 0$. Since $\rho \sim a^{-4}$ 
decreases with $a$ more rapidly then $\rho_{k} \sim a^{-3}$ (or even $\sim a^{-2}$ 
\cite{KK2}), the era of accelerating expansion should begin with increasing $a$, even if 
the state with $\rho_{k} < \frac{2}{3} \,\rho$ and $\Lambda \sim 0$ 
existed in the past, when the expansion was decelerating. A condensate of quantized 
primordial scalar field can be identified with a dark energy \cite{KK2, KK3}.

Let us calculate the corrections $\rho_{u}$ and $\widetilde{\rho}_{u}$. These terms are 
essential in the very early universe at $a < 1$ \footnote{For the present-day 
Universe we have $a \sim 10^{61}$ in accepted dimensionless units.}. This quantum theory
predicts the quantum origin (nucleation) of the universe from the region  $a \sim 0$
\cite{KK1}. It means that the state vector in this region is constant, 
$\langle a \sim 0| f_{k}\rangle = \mbox{const}$. For such a state
\begin{equation}\label{45}
    S = \frac{i}{2}\, \ln a + \mbox{const}
\end{equation}
and the quantum corrections (\ref{40}) and (\ref{41}) are equal to
\begin{equation}\label{46}
    \rho_{u} = - \frac{1}{4a^{6}}, \qquad 
    \widetilde{\rho}_{u} = -\frac{i \dot{a}}{2a^{4}} = - \frac{i}{2 a^{5}}\,
        \partial_{a}S = - \frac{1}{4a^{6}},
\end{equation}
where we used the representation (\ref{21}) for the calculation of 
$\widetilde{\rho}_{u}$.
It can be done in the semi-classical approach under consideration\footnote{Let us note 
that the presence of a minus sign in $\rho_{u}$ (\ref{46}) is not
extraordinary. According to quantum field theory, for instance, vacuum fluctuations
make a negative contribution to the field energy per unit area (the Casimir effect).}. 

With account of equations (\ref{46}),
Eqs (\ref{38}) and (\ref{39}) can be reduced to the form of the standard Einstein equations for the homogeneous, isotropic and closed universe
\begin{eqnarray}\label{47}
    \left(\frac{\dot{a}}{a}\right)^{2} = \rho_{tot} - \frac{1}{a^{2}}, \qquad \frac{\ddot{a}}{a} = -\,\frac{1}{2}\,\left[\rho_{tot} + 3 p_{tot}\right],
\end{eqnarray}
where the quantities
\begin{equation}\label{48}
  \rho_{tot} =  \rho_{m} + \rho_{u}, \qquad  p_{tot} = p_{m} + p_{u}
\end{equation}
describe the total energy density and the pressure of the matter in the universe which
take into account its quantum nature in semi-classical approximation.
The quantum correction $\rho_{u}$ may be identified with the ultrastiff matter with the equation of state
\begin{equation}\label{49}
    p_{u} = \rho_{u},
\end{equation}
where $p_{u}$ is the pressure. This `matter' has quantum origin.

Let us estimate the ratio of energy density $|\rho_{u}|$ to $\rho_{m}$. Passing
to the ordinary units, we have
\begin{equation}\label{50}
    \mathcal {R} \equiv 
    \left[\left(\frac{2 G \hbar}{3\pi c^{3}}\right)^{2} \frac{1}{4 a^{6}}\right]: 
     \left[ \frac{8\pi G}{3c^{4}}\,\rho_{m} \right] .
\end{equation}
where $\rho_{m}$ is measured in GeV/cm$^{3}$ and $a$ in cm. 
For our Universe today $\rho_{m} \sim 10^{-5}$ GeV/cm$^{3}$, $a \sim 10^{28}$ cm and
\begin{equation}\label{51}
    \mathcal {R}_{today} \sim 10^{-244},
\end{equation}
i.e. the quantum correction may be neglected to an accuracy of $\sim O(10^{-244})$. 
In the Planck era $\rho_{m} \sim 10^{117}$ GeV/cm$^{3}$, $a \sim 10^{-33}$ cm and the relation 
\begin{equation}\label{52}
    \mathcal {R}_{Planck} \sim 1
\end{equation}
shows that the densities $\rho_{m}$ and $\rho_{u}$ are of the same order of magnitude.

\textbf{4.2. Quantum effects on sub-Planck scales.} On sub-Planck scales, $a< 1$, the
contributions from the condensate, cosmological constant 
and curvature may be neglected. As a result the equations of the model take the form
\begin{equation}\label{53}
    \frac{1}{2}\,\dot{a}^{2} + U(a) = 0,\qquad  \ddot{a} = - \frac{dU}{da},
\end{equation}
where
\begin{equation}\label{54}
    U(a) \equiv \frac{1}{2}\, \left[ \frac{1}{4 a^{4}} - \frac{E}{a^{2}} \right].
\end{equation}
These equations are similar to ones of Newtonian mechanics. Using this analogy
they can be considered as equations which describe the motion of a `particle'
with a unit mass and zero total energy under the action of the force
$-\frac{dU}{da}$, $U(a)$ is the potential energy, and $a(\tau)$ is a generalized 
variable.
A point $a_{c} = \frac{1}{2 \sqrt{E}}$, where $U(a_{c}) = 0$, divides the region of 
motion of a `particle' into the subregion $a < a_{c}$, where the classical motion of 
a `particle' is forbidden, and the subregion  $a > a_{c}$, where the classical
trajectory of a `particle' moving in real time $\tau$ exists.

In the subregion $a < a_{c}$ there exists the classical trajectory of a `particle' 
moving in imaginary time $t = - i \tau + \mbox{const}.$ in the potential $- U(a)$.
Denoting the corresponding solution as $\tilde{a}$ we find
\begin{equation}\label{55}
   \tilde{a} = a_{c} \sin z, \qquad 
       t = \frac{a_{c}^{3}}{2}\, [2 z - \sin 2z].
\end{equation}
At small $z$, i.e. in the region $\tilde{a} \sim 0$, we have
\begin{equation}\label{56}
    \tilde{a} = \left(\frac{3}{2}\,t \right)^{1/3}.
\end{equation}
Comparing Eq. (\ref{56}) with the standard model solution  (see, e.g. \cite{KolT, DF}), 
we conclude that it agrees with the fact that the
`matter' near the point $\tilde{a} = 0$ is described by the equation of state of the
 ultrastiff matter (\ref{49}).

\begin{figure}[ht]
\begin{center}
\subfigure[$E = 1$]{\label{fig:1-a}\includegraphics*[scale=0.6]{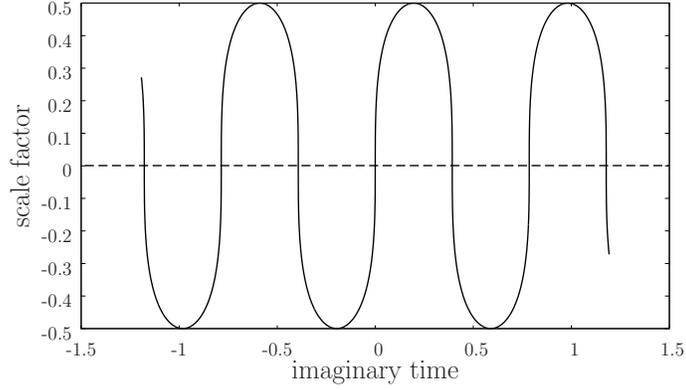}}
\subfigure[$E = 3$]{\label{fig:1-b}\includegraphics*[scale=0.6]{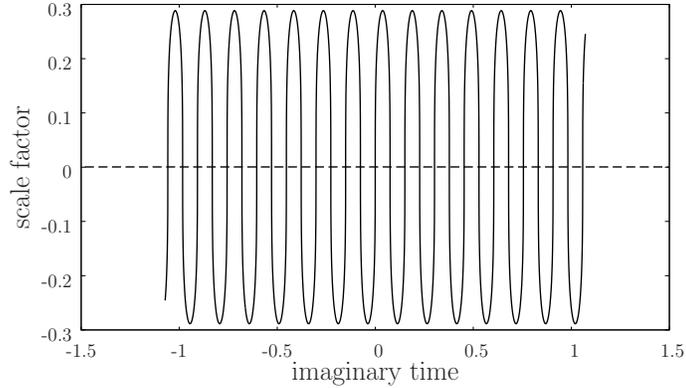}}
\end{center}
\caption{The scale factor $\tilde{a}$ vs. imaginary time $t$ for the cases
$E = 1$, which corresponds to the ground state $n = 0$ (a), and
$E = 3$ for the state with $n = 1$ (b). 
} \label{fig:1}
\end{figure}

In Fig.~1 the scale factor $\tilde{a}$ is shown as a function of imaginary time $t$. It 
demonstrates that the scale factor oscillates in imaginary time near its zero value. The
amplitude and frequency of these oscillations depend on the parameter $E$. 
In the subregion $a \leq a_{c}$, where the contributions from the condensate and 
cosmological constant may be neglected, the eigenvalue $E$ of Eq. (\ref{29}) is
quantized according to the expression (\ref{31}) with $M_{k} = 0$. The case
$E = 1$ corresponds to the universe which is in the ground state $n = 0$ (see Fig.~1a).
The value $E = 3$ refers to the quantum state $n = 1$ (see Fig.~1b). 

Since the equations (\ref{53}) are invariant with respect to a substitution
$a \rightarrow -a$, as well as the metric (\ref{1}) is, the existence of negative
values of $a$ with $|a| \leq 0.5$ does not contradict with ansatz. On the other hand,
it means that from the point of view of quantum description the sign of $a$ appears to be
indefinite on sub-Planck scales. The states of the universe $\langle a|f_{k} \rangle$
in that era ($M_{k} = 0$) separate into even and odd states with respect to the sign
of $a$, $\langle - a|f_{k} \rangle = (-1)^{n} \langle a|f_{k} \rangle$. Such a symmetry
will be broken when a condensate of the scalar field with the mass
$M_{k} > 0$ begins to influence the dynamics of the universe as a whole. In the region
$a \apprge 1$, where the universe should still be considered as a quantum system
with the equation of motion (\ref{29}), the mean value of the scale factor in the
state $\langle a|f_{k} \rangle$ is equal to 
$\langle f_{k}|a|f_{k} \rangle = M_{k} + O\left ((2 M_{k})^{2n - 1} e^{-M_{k}^{2}}) \right )$,
where $\langle f_{k}|f_{k} \rangle = 1$ \cite{KK1}, and the scale factor in 
the equations (\ref{47}), which corresponds to this mean value, takes the positive 
values only. In fact, in the semi-classical approach under consideration the
choice of a sign of the scale factor is realized even earlier under the transition
of the universe from the subregion of a motion in imaginary time into the subregion
of a motion in real time, $a > a_{c}$ (see Fig.~2 below).

The amplitude of oscillations which determines the size of the region forbidden for 
the classical motion decreases as $n^{-1/2}$, while the frequency increases 
almost as $n^{3/2}$ with 
the increase of $n$. The case of very high values of $n$ describes the classical motion
of the system. In the limit $n \rightarrow \infty$ the region forbidden for 
the classical motion shrinks to the point producing the initial singularity in which
the universe is characterized by the infinitely large frequency of oscillations of space
curvature.

In the subregion $a > a_{c}$ the solution of the equations (\ref{53}) 
can be written as
\begin{equation}\label{57}
    a = a_{c}\cosh \zeta, \qquad 
     \tau = \frac{a_{c}^{3}}{2}\, [2 \zeta + \sinh 2\zeta].
\end{equation}
At $\zeta \ll 1$ it follows from here that
the scale factor at $\tau \ll 2 a_{c}^{3}$ increases almost exponentially
\begin{equation}\label{58}
    a = a_{c}\left[1 + \left(\frac{1}{2a_{c}^{2}}\right)^{3} \tau^{2} + \ldots \right]
       \approx a_{c}\, \exp \left\{\frac{\tau^{2}}{8 a_{c}^{6}} \right\}. 
\end{equation}

The almost exponential expansion of the early universe in that era is
stipulated by the action of quantum effects which, according to Eqs (\ref{46})
and (\ref{49}), cause the negative pressure, $p_{u} < 0$, i.e. produce an 
anti-gravitating effect on the cosmological system under consideration.

At $\zeta \gg 1$ the solution (\ref{57}) takes the form
\begin{equation}\label{59}
    a = \left(\frac{\tau }{a_{c}}\right)^{1/2}.
\end{equation}
It describes the radiation dominated era and corresponds to time
$\tau \gg 2a_{c}^{3}$.

\begin{figure}[ht]
\begin{center}
\subfigure[$n = 0$]{\label{fig:2-a}\includegraphics*[scale=0.6]{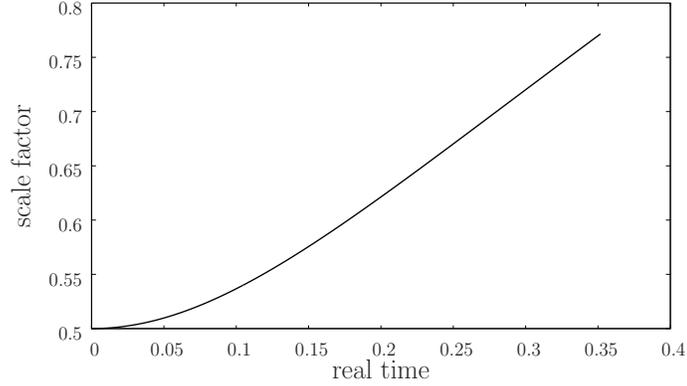}}
\subfigure[$n = 1$]{\label{fig:2-b}\includegraphics*[scale=0.6]{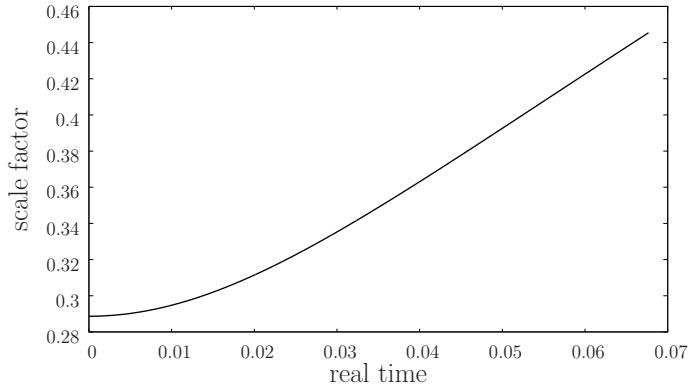}}
\end{center}
\caption{The scale factor $a$ vs. real time $\tau$ for the values $n = 0$ (a),
and $n = 1$ (b). 
} \label{fig:2}
\end{figure}

In Fig.~2 the scale factor $a$ is shown as a function of real time $\tau$. With the
increase of $\tau$, it increases at first by the law (\ref{58}), and then in accordance 
with Eq. (\ref{59}). The initial value $a(\tau = 0)$ depends on the quantum number $n$.
Fig.~2a demonstrates the case with $n = 0$, while Fig.~2b shows the case of $n = 1$. 
In the limit $n \rightarrow \infty$ the initial singularity $a(\tau = 0) = 0$ 
will be the reference point of the scale factor $a$ as a function of $\tau$ 
as in general relativity.

The solutions (\ref{55}) and (\ref{57}) are related between 
themselves through an analytic continuation into the region of complex values
of the time variable,
\begin{equation}\label{60}
    t = - i \tau + \frac{\pi}{2}\,a_{c}^{3}, \qquad z = \frac{\pi}{2} - i \zeta.
\end{equation}
The scale factors $\tilde{a}$ (\ref{55}) and $a$ (\ref{57}) are connected through the 
condition
\begin{equation}\label{61}
    a(\tau) = \tilde{a} \left(\frac{\pi}{2}\,a_{c}^{3} - i \tau \right),
\end{equation}
which describes an analytic continuation of the time variable $\tau$ into the region
of complex values of Euclidean time $t$. This analytic continuation may be
interpreted as a quantum tunneling of the Lorentzian space-time from the Euclidean one.

\textbf{4.3. Transition amplitude.} 
The model determined by the equations (\ref{53}) allows us to describe the origin 
(nucleation) of the universe as the transition from the state in the subregion
$a < a_{c}$ to the state in the subregion $a > a_{c}$. The corresponding 
transition amplitude can be written as follows \cite{Col}
\begin{equation}\label{62}
    T \sim e^{-S_{t}},
\end{equation}
where $S_{t}$ is the action on a trajectory in imaginary time $t$,
\begin{equation}\label{63}
    S_{t} = 2 \int_{-\infty}^{\infty}\!\!dt U(\tilde{a}).
\end{equation}
Let us proceed to the integration with respect to the time variable $z$. According
to Eq. (\ref{55}) the scale factor $\tilde{a}$ is a periodical function of $z$.
We shall at first consider the oscillations of a `particle' on the finite time interval 
$[- z_{0}, z_{0}]$ with the boundary conditions $\tilde{a}(z_{0}) = \pm a_{c}$ 
and $\tilde{a}(- z_{0}) = \mp a_{c}$. Supposing that $z_{0} = \frac{\pi}{2} \nu$,
where $\nu = 1,3,5,\dots$ numbers the quantity of half-waves of the function
$\tilde{a}(z)$, centered at the points $z = \pm \pi q$, $q = 0,1,2,\dots$,
which cover the interval $[- z_{0}, z_{0}]$. Then the action $S_{t}$
takes the form
\begin{equation}\label{64}
    S_{t} = 2 \int_{-\frac{\pi}{2} \nu}^{\frac{\pi}{2} \nu}\!\!
       dz \frac{dt}{dz} U(\tilde{a}(z)).
\end{equation}
Using the explicit form of the solution (\ref{55}) we find
\begin{equation}\label{65}
    S_{t} = - \sqrt{E} \pi \nu,
\end{equation}
and the amplitude (\ref{62}) becomes
\begin{equation}\label{66}
    T \sim e^{\sqrt{E} \pi \nu},
\end{equation}
i.e. a `particle' which is the equivalent of the universe leaves the subregion forbidden
for classical motion with an exponential probability density. It is pushed out of 
forbidden subregion into the subregion of very small values of $a$ in real time $\tau$
by the anti-gravitating forces stipulated by the negative pressure which cause 
quantum processes at $a \sim 0$ (see Eqs (\ref{46}) and (\ref{49})). This phenomenon
can be interpreted as the origin of the universe from the region $a < a_{c}$. It is
possible only if the probability density that the universe is in the state with
$a \sim 0$ is nonzero.

In the 
limit $\nu \rightarrow \infty$ the transition amplitude $T \rightarrow e^{\infty}$.
This result may be interpreted so that the origin of the universe occurs with necessity
during the infinite imaginary time interval.

\textbf{4.4. Geometry.} Let us consider how the geometry of the universe changes as a
result of its 
transition from the region $a < a_{c}$ into $a > a_{c}$.
In the model under consideration the metric has the form (\ref{1}).
According to the solutions (\ref{55}) and (\ref{57}) the
metric (\ref{1}) takes the form
\begin{equation}\label{67}
    ds_{E}^{2} = - a_{c}^{2}\,\sin^{2} z \left\{4 a_{c}^{4} \sin^{2} z\,
        dz^{2} + d\Omega_{3}^{2} \right\} \quad \mbox{at} \quad a < a_{c}
\end{equation}
and
\begin{equation}\label{68}
    ds_{L}^{2} = a_{c}^{2}\,\cosh^{2} \zeta \left\{4 a_{c}^{4} \cosh^{2} \zeta\,
        d\zeta^{2} - d\Omega_{3}^{2} \right\} 
        \quad \mbox{at} \quad a > a_{c},
\end{equation}
where the interval with the Euclidean signature is denoted by the index $E$, and the one
with the Lorentzian signature is marked by $L$. Introducing the new time variables $\xi$
and $\varsigma$ according to
\begin{equation}\label{69}
    d\xi = 2 a_{c}^{2} \sin z dz, 
        \qquad d\varsigma = 2 a_{c}^{2} \cosh \zeta d\zeta,
\end{equation}
the metrics (\ref{53}) and (\ref{54}) can be reduced to the conformally flat form
\begin{equation}\label{70}
    ds_{E}^{2} = - a_{c}^{2} \left[1 - \left(\frac{\xi}{2 a_{c}^{2}}\right)^{2} \right]
        \left\{d\xi^{2} + d\Omega_{3}^{2} \right\},
\end{equation}
\begin{equation}\label{71}
    ds_{L}^{2} = a_{c}^{2} \left[1 + \left(\frac{\varsigma}{2 a_{c}^{2}}\right)^{2} 
        \right] \left\{d\varsigma^{2} - d\Omega_{3}^{2} \right\}.
\end{equation}
Both metrics are related between themselves through the analytic continuation
into the region of complex values of the time variable $\varsigma = i \xi$.
The conformal factor in the metric (\ref{70}) varies from zero value at 
$\xi = -2 a_{c}^{2}$
to the maximum value $a_{c}^{2}$ at $\xi = 0$, and then vanishes again at 
$\xi = 2 a_{c}^{2}$. The conformal factor in the metric (\ref{71}) ranges from
its minimum value $a_{c}^{2}$ at $\varsigma = 0$ to infinity increasing 
with the increase of $\varsigma$.

The metric (\ref{70}) is conformal to a metric of a unit four-sphere in a 
five-dimensional Euclidean flat space. With increasing $a$, the universe transits
from the region $a < a_{c}$ into the region $a > a_{c}$,
where the geometry is conformal to a unit hyperboloid embedded in a 
five-dimensional Lorentz-signatured flat space.
Such a picture of change 
in spacetime geometry during the transition of the universe from the
region near initial singularity into the region of real physical scales agrees with the
hypothesis \cite{HH,HP}, widely discussed in the literature (see, e.g., the reviews 
\cite{AB,KS}) for the de Sitter space, about possible change in four-space 
geometry after the spontaneous nucleation of the expanding universe from the initial
singularity point.

\begin{center}
      \textbf{5. Concluding remarks }\\[0.3cm]
\end{center}

In this paper we study the properties of the quantum universe on extremely small 
spacetime scales in the semi-classical approach to the well-defined quantum model. 
We show that near the initial cosmological singularity point quantum gravity effects
$\sim \hbar$ exhibit themselves in the form of additional matter source with the 
negative pressure and the equation of state as for ultrastiff matter. The analytical
solution of the equations of theory of gravity, in which matter is represented by
the radiation and additional matter source of quantum nature, is found. It is shown
that in the stage of the evolution of the universe, 
when quantum corrections $\sim \hbar$
dominate over the radiation, the geometry of the universe is described by the
metric which is conformal to a metric of a unit four-sphere in a five-dimensional 
Euclidean flat space. In the radiation dominated era the metric is found to be
conformal to a unit hyperboloid embedded in a five-dimensional Lorentz-signatured flat
space. One solution can be continued analytically into another.

The origin of the universe can be interpreted as a quantum transition of the
system from the region in a phase space forbidden for classical motion, but where
a trajectory in imaginary time exists, into the region, where the equations of motion
have the solution which describes the evolution of the universe in real time. Near the
boundary between two regions, from the side of real time, the universe undergoes
almost an exponential expansion which passes smoothly into the expansion under the
action of radiation dominating over matter which is described by the standard
cosmological model. As a result of such a quantum transition the geometry of the
universe changes. This agrees with the hypothesis about the possible change
of geometry after the origin of expanding universe from the region near the initial 
singularity point.
In this paper this phenomenon is demonstrated in the case of the early universe
filled with the radiation and ultrastiff matter which effectively takes into account
quantum effects on extremely small spacetime scales.
The properties of the universe on sub-Planck scales 
do not depend on the form of primordial matter (the scalar field in the model under 
consideration) and one can conclude that they are model-independent in this sense.
We describe the mechanism of a shrinkage of the region forbidden for the classical motion
to the point of the initial cosmological singularity.

\end{document}